# Non-specular Reflective Optics

By

Timir Datta,
Ming Yin[1],
Yunjin Wang, Michael Wescott, Richard Foster and Rebecca Bowers

Physics and Astronomy Department, University of South Carolina, Columbia, SC 29208
1. Physics and Engineering Department, Benedict College, Columbia, SC 29204

**Overview:**

Geometrically decorated two-dimensional (2D) discrete surfaces can be more effective than conventional smooth reflectors in managing wave radiation. Constructive non-specular wave scattering permits the scattering angle ($\beta$) to be other than twice that of incidence ($\alpha$). This results in gross violations of the law of reflection so that significant fraction of the $\alpha$, $\beta$ phase space becomes accessible. A wide range of novel reflective behaviors ensues; including the phenomenon of negative reflection were energy transport remains on the same side of the normal. Also, at a critical incidence [$\alpha_{crit} = \cos^{-1}(1-\kappa)$ or $\cos^{-1}(\kappa-1)$] coherent superposition can force both the transmitted and reflected waves to graze the scattering surface ($\alpha = \beta$) thus synergistically reinforcing the diffractive process in a behavior reminiscent of critical internal reflection of ray optics.

We experimentally demonstrate the concept with measurements on a one-dimensionally periodic system (grating) where the relation $\alpha = \cos^{-1}[Z(\beta, \kappa)]$ holds; Z is shown to be a function of the diffractive index parameter ($\kappa$) and the two angles $\alpha$ and $\beta$. Excellent agreement is found between experimental data and theory. Future applications include generalized zone plate optics, motionless heliostats, beam splitting, evasive and deceptive imaging and high precision metrology. The relevant 2D-surfaces or meta-materials (2DMM) require patterns equal to less than a few wavelengths deep hence are far convenient than their 3-dimensional counterparts. Furthermore the technology to fabricate 2DMM structures with functionalities even in optical frequencies will be much less daunting. And, this potential for on-demand, adaptive maneuverability of optical properties that is most attractive.

**Background**:

In high performance optics attention is lavished in producing compositionally homogenous, bubble and stress free materials[1-3]. For reflective purposes, mirrors follow ray or geometrical principles and are amongst the simplest and most ancient "optical" instruments. However a recent trend is towards the so-called metamaterials and related systems[4-12]. This progression from homogenous perfection to finely crafted structuring is an evolutionary process akin to that in electronics. The motivation for both has been to take full advantage of the wave nature of electrons[13] and photons by precise control of superposition.

Geometric arrangements such as superlattices[14] and quantum wells that optimize the energy dispersion of the quantum nature of the electron made a great impact in electronics. Likewise, ordered composites or metameterials have given rise to localized or stopped light [15, 16], violation of Snell's law and negative refraction [8-12]. Phenomena that may appear counter intuitive but are created by coherent superposition or reemission[17] of waves from an array of meso-scale embellishments tailored into the system. These meticulously engineered features promote highly directional coherent scattering to constructively interfere to hold back transport or spatially restrict the radiated energy.

We reason that a modular reflecting system that takes advantage of the cyclic nature of radiation can be far more versatile in directing energy than traditional smooth reflectors. Gratings and echelons are 1-dimensionally (1D) modular systems that are well known for spectral resolution, but are not optimized for the management of reflection which is the principle concern of this article. Over the entire range of incident angle ($\alpha$) the scattering angle ($\beta$) sweeps through the full circle to cover a large portion of the $\alpha$, $\beta$ phase plane and energy distribution can be dominated by either positive or negative reflection. In the later case the reflected wave is back scattered and emerges on the same side of the normal as the incident wave so most of the incoming energy return to the same medium via a negative angle. Under suitable conditions, a pathological diffractive phenomenon reminiscent of critical internal reflection can result, in this case coherent superposition forces both the transmitted and reflected waves to be degenerate and graze the scattering surface thus synergistically reinforcing the cumulative process all over again.

From the aspect of engineering feasibility, many of the currently available techniques for 3D optical metameterials are well suited for exploring the 2D structures described here. The technology for fabricating the 2D structures over a wide range of frequencies ranging from infrared to radio waves, even those with smart scattering properties is achievable.

**Coherent non-specular reflection:**

The physical description and associated mathematics depends on the wave versus the particle points of view. With wave the discussions relates to "optical paths" ($\Delta x$) and phase changes ($\Delta \phi$) but the particle descriptions concern with linear momentum **p(k)** and related conservation principles such as $\Delta \mathbf{k} = \mathbf{G}$[13]. Here we take the wave description and explicitly discuss electromagnetic radiation[18,19] although similar considerations hold for other forms of waves as well.

To understand the influence of meso-scale regularities on scattering let us consider a wave incident along AO strikes a boundary between two media (figure 1). At O the waves may constructively superpose along the four directions indicated. Two, OB and OB' are reflected hence remain in the first medium, OC and OC' propagate and refracted in the second medium. In a perfectly translation invariant, uniform and structureless systems energy propagate forward because back flow of waves are suppressed by Huygen-Fresnel[2-3] principle. Hence, at a planar interface the waves may exclusively undergo specular scattering and the radiation falling back into the first medium takes a corpuscular trajectory. In figure 1 only two "rays" (OB and OC) are consistent with geometric optics. OB being the most banal and satisfy the corpuscular theory, the refracted ray OC was the nemesis and the clincher for the eventual acceptance of the wave theory or physical optics.

Fortunately, suitably periodic scatterings can be tailored to break this principle. Violation of the laws of reflection (and refraction) requires the interaction to introduce a negative wave number (momentum) parallel to the interface. The other two paths in figure 1, namely OB' and OC' both make negative angles as they remain on the same side of the normal and are totally non-specular.

Figure 2 provides an enlarged diagram of the scattering process. Unlike traditional optical convention here a more practical coordinate system fixed to the laboratory is chosen; the symmetry breaking wave vector ($k_1$) of the incoming radiation defines the x-axis and the angles are measured counter clock wise with respect to positive-x. The periodicity vector, $d$ is also shown. The incident direction or attitude ($0°<\alpha<180°$) of the line of intersection between the scattering surface and the plan of the figure and the angle ($\alpha<\beta<360°$) of the out going wave are indicated. The scattering surface need not be a simple one-dimensional line grating or a lattice of spots but in principle it can be decorated with any regular motif, even of tessellated "girih" arrangement of Archimedes or Penrose tiles[20, 21] in which case diffraction will be quite interesting. Furthermore the markings and their spacing can be dynamically controllable.

The nth order constructive superposition condition is as follows

$$\Delta\phi_{total} = \Delta\phi_1 + \Delta\phi_2 = n \bullet 2\pi \qquad \ldots 2.1$$

Where only the absolute values are relevant also the phase differences are related to the path differences, $\Delta x_i$ by

$$\Delta\phi_i = k_i \bullet \Delta x_i$$

so that in terms of the periodicity vector $d$, we have

$$\Delta\phi_{total} = \Delta k \bullet d$$

or,

$$\Delta\phi_{total} = k_2 d \cdot Cos\theta_2 - k_1 d \cdot Cos\theta_1 \qquad \ldots 2.2\ a$$

Where, $\theta_1$ and $\theta_2$ are the angles between the vector $d$ and the incoming and out going wave vectors respectively. Notice that in the present case the incident and out going angles are not forced to be equal. Assuming the medium to be the same everywhere and for specificity let the surface be one dimensionally periodic[20] with a single grating constant d, then equation 2.1 may be re-expressed in terms of the quantities shown in figure 2, as:

$$Cos(\beta - \alpha) - Cos(\alpha) = \kappa \qquad \ldots 2.2\ b$$

Where, the diffractive index is given by

$$\kappa = n\frac{\lambda}{d}$$

Equations 2.2 are the central relations for scattering of a wave from a periodic surface. These equations are not mere 2D versions of the familiar Bragg law[18] $2k\sin\theta = G$ but they impose circular functional dependence of between the angles ($\alpha$ and $\beta$) so that each one may be determined when the value of the other angle is known. For instance,

$$\alpha = Cos^{-1}[Z(\beta,\kappa)_1] \quad \ldots 2.3a$$

or,

$$\beta = Cos^{-1}[Z(\alpha,\kappa)_2] \quad \ldots 2.3b$$

where,

$$Z(\beta,\kappa)_1 = \frac{\kappa}{2}[1 \pm \sqrt{\{1 - \frac{2(1-\frac{Sin^2\beta}{\kappa^2})}{(1-Cos\beta)}\}}]$$

and

$$Z(\alpha,\kappa)_2 = [Cos\alpha \bullet (\kappa + Cos\alpha) \pm |Sin\alpha|\sqrt{\{1-(Cos\alpha+\kappa)^2\}}]$$

Equations 2.3 indicate a non-linear relationship between the three quantities, namely the two angles ($\alpha$ and $\beta$) and the parameter $\kappa$. In this case of 1D periodicity the influence of the surface, wavelength and the diffraction order manifests only via this single diffraction index parameter ($\kappa$).

Quite generally, $0 < \alpha < \pi$; and the solutions span all possible scatterings i.e., both reflection and transmission. However the behavior of the transmitted radiation will be discussed elsewhere[20] and presently we will be concerned with only the reflective process. The dashed diagonal red line ($\beta = 2\alpha$) in figure 3 is the well-known specular or law of reflection solution. This line is the seperatrix and covers only a set of measure zero of the $\alpha$, $\beta$ plane. The other solutions represent the violation of the law of reflection, and comprise of positive and negative reflections occupying a total area equal to half of the plane. The totality of reflective solutions occupies half of the $\alpha$, $\beta$ space, positive (two diagonal regions, grey) and negative solutions (two off- diagonal regions, light grey) covering equal areas (figure 3). In the regime ($0<\alpha<\pi/2$), $\beta$ for negative reflection is bounded by $\alpha< \beta< \alpha+\pi$ and for $\pi/2 <\alpha < \pi$ the $\beta$ has to be $\alpha< \beta< \alpha+\pi/2$.

In the $\alpha$, $\beta$ space, as $\kappa$ ranges from 0 to 2, the two critical lines, $\beta = \alpha$ and $\beta= \alpha + \pi$ define the boundaries between transmission (but no reflection) and the "reflection permitted" regions (no transmission allowed). They represent "pathological diffraction" that is evocative of critical internal reflection of ordinary optics. Physically constructive superposition process forces the emergent reflected and transmitted "rays" to collapse and graze the scattering interface. This synergistic critical scattering, resonantly increases the inter dependence between reflection and transmission. From equation 2.2b the corresponding solutions are as follows,

$$\alpha_{critical} = Cos^{-1}(1-\kappa) \& Cos^{-1}(\kappa-1)$$

Hence for: $\kappa = 0$, $\beta = \alpha_{crit} = 0$ & $\beta = 180°$; $\kappa = 1$, $\beta = \alpha_{crit} = 90$ & $\beta = 270°$ and $\kappa = 2$, $\beta = \alpha_{crit} = 180$ & $\beta = 360°$ respectively. We have tested these results with experiments at several wave lengths in the red to green region of the visible spectrum using a ruled gold film as the grating. A demonstration of the positive vis-a-vis negative reflection of a red and a green laser beams from the patterned (gold) surface is provided in figure 4. Here the left panel shows the red laser entering the setup at the top right corner goes straight through a semi-silvered beam splitter and after passing through an opening in the screen is incident on the gold surface (shown in silhouette). The green beam enters from the left but is rendered parallel to the red beam by the beam splitter. Upon scattering, the positively reflected beams move to the left of the normal and the negative reflections remain on the same (right) side of the normal as the incident beams. The sensitivity of the negative reflection is illustrated in the right panel of figure by the large angular separation between red and green. The two beams were vertically offset so that the spots on the screen will not completely overlap.

Graphs of the calculated $\alpha$ and $\beta$ for six values of $\kappa$ (0.4, 0.6, 0.8, 1.0 & 1.2) are shown in figure 5. Also included are two sets of experimental data obtained for the red and green lasers. Excellent agreement between experiment and theory is obtained. As anticipated equations 2.2, over the range of out going angles, the graphs clearly display a non-trivial dependency. The solid vertical (orange) line indicates that for $\kappa$ equal to 0.6 (blue curve) at $75^0$ incidence there is one positive reflection (marked a) at about $110^0$ and a negative reflection (marked b) at about $190^0$. Also note for $\kappa$ equal to 1.0 and at $90^0$ incidence how abruptly scattering switches over from negative to positive. Over certain regimes the exiting angle is insensitive to the scattering angle. The stationary regions are of interest in spectrometer design to insure that misalignments of the grating do not severely poison the wavelength determination. On the other hand as indicated earlier, in the interfaces demarcated by the two lines, $\beta = \alpha$ and $\beta = \alpha + \pi$ critical diffraction forces both the "emergent rays" to collapse and graze the interface that synergistically enhances scattering. Hence, at slightly higher (lower) incidences the rotational sensitivity is greatly heightened so that a minimal change in $\alpha$ greatly affects the exiting angle $\beta$.

**Technological implications:**

In the previous sections we have described how modular surfaces produce a far complex reflective phenomenon compared to traditional smoothly polished mirrors. Quantitatively, this richness was presented as the difference in the coverage of the $\alpha$, $\beta$ phase space accessible to coherent specular versus non-specular scatterings. We reason that these present exciting scientific and technological opportunities. In the following sections we outline several potential technological implications.

As evident in figure 4 under suitable conditions a single incident beam can be rendered into multiple beams; furthermore if desired the relative intensity ratios can be varied by means of $\alpha$ and $\kappa$. In the familiar semi-silvered (half-reflection) splitter one of the beams is reflected and the other is transmitted, hence the beams have different histories and unequal phase sifts. As a consequence, conventional splitters are bulky because the reflected beam requires additional compensating optics. In contrast in our process the beams emerge into the same medium and with same optical history because they are all reflected. Currently in the telecommunication industry, in photonic computation and in quantum information related technology there are wide spread application of beam splitters. We expect beam splitting described here is better suited particularly in monochromatic applications.

Second, suitably decorated 2D surfaces are generalized zone plates thus can readily transform an incident plane wave into a focused spherical wave. Hence such a surface with smart attributes can function as planar imaging elements with adaptive properties. Furthermore this type of devices will reduce the physio-chemical complexities associated with manufacturing 3D optical materials to the problem of 2D lithography.

Next we will outline a "heliostat", where the Golan is to hold constant $\beta$, even when $\alpha$ changes with time. In conventional heliostats the guiding optical elements are bodily moved or rotated, but in our present case $\beta$ can be rendered stationary by varying the parameter $\kappa$ (or d) with out any physical movement! Such devices can be of use in energy control to hold the spot at a desired constant position focused on a receiving satellite, photovoltaic converter or a sweet spot of a boiler. Also different parts of the solar spectrum can be simultaneously directed to a number of cells each optimized for best conversion efficiency. Similar systems are also of interest in radar applications for evasive purposes as well as for masking the size, shape and speed of the target. At micro-wave frequencies such variable geometry structures can be mechanical, electromechanical as well as totally electronic. Figure 6 shows the calculated values of d/$\lambda$ as the incident angle ranges between 40°-140° for six constant values of $\beta$.

Detection of rotation and angle measurement is quite important. Currently a wide variety of methods for rotation measurements are extant, for example interference techniques are common, auto-collimation is also used and many others are described both in the scientific and in the patent literature [3,22-25]. In the visible regime the "mirror and scale" is the most familiar and simplest example of such techniques. As the mirror turns by $\delta\omega$ degrees the reflected ray is rotated by the angle $\Delta\Omega$; the ray optics method comes with an automatic two fold magnification factor i.e., the outgoing ray under goes deflection $\Delta\Omega$ given by

$$\Delta\Omega = 2\delta\omega \qquad \ldots \quad 3.1$$

Since the advent of solid state lasers and detectors a number of variations of this reflection technique have become the method of choice even in such demanding applications such as for monitoring the stylus in atomic force microscopes.

Enhanced sensitivity of $\beta$ to small changes in $\alpha$ is another motivation to harness non-specular scattering for the purpose of detecting and measuring angular displacements. Over certain regions of the phase space the scattered spot is very sensitive to orientation, our experimental results agree quite well with the calculated results. For example, in the regime of interest for rotation detection, a twist of 0.07 deg of the scattering surface the spot is deflected by more than ten times. Amplification is a sensitive function of the incident wavelength and angle, as can be seen from equation 2.2. And over limited angles still higher gains are conceivable.

As another application consider radiations of different wavelengths. The corresponding $\beta$ angles and the distance $\Delta x$ between the reflected spots is given by

$$\Delta x = R \bullet [\beta(\lambda_1) - \beta(\lambda_2)] = R \bullet \Delta\beta \qquad \ldots \quad 3.1$$

From equation 3.1 the value of *R,* the distance to the target can be determined from $\Delta x$ and the theoretical values of $\beta$, as follows

$$R = \frac{\Delta x}{\beta(\lambda)_1 - \beta(\lambda)_2} \qquad \ldots 3.2$$

Notice, in equation 3.2 the unknown distance R is being determined from the known values of β and the wave lengths. The confidence level in R can be further improved by increasing the number of wavelengths employed. Likewise, using a monochromatic radiation but with multiply periodic reflector one can also determine the target distance, from the following relation ship,

$$R = \frac{\Delta x}{\beta(d_1) - \beta(d_2)} \qquad \ldots 3.3$$

Such schemes are not possible with conventional smooth reflection because there is no distinguishing feature between one smooth flat surface from another so the reflected angle is independent of wave length. Also this distance measurement is not triangulation or time of flight because it does not entail angle nor time measurements; furthermore, as indicated earlier, negative reflection allows a choice of angels and wave lengths to obtain the "best" spot separation Δx.

**Summary**:

Seeking violations of established principles is an effective means to improve both the understanding and the application of natural principles. Here we have shown that it is possible to violate the law reflection and obtain novel phenomena such as negative reflection which in many respects is similar to negative refraction. A reflecting system that takes advantage of the wave nature of light will be far more versatile than smooth reflectors. The detailed nature of the patterned surface is application dependant. For simple redirection of the incident energy in the optical regime a simple one-dimensional reflection grating will be adequate. We also outlined a number of potential applications. Generalized zone plates for imaging with zero-curvature focal length control, novel transmission free beam splitting, and motion less heliostats are such applications. Another example is the on-demand control of radar cross section and enhancement of angular sensitivity so that compared to the factor two obtained by "mirror and scale" a single negative-reflection-pass can provide a factor of at least ten in angle amplification, even higher gains are possible. We demonstrated the feasibility of non-specular scattering with a simple 1D diffractive system. It is shown that our experimentally observed behaviors are well described by conventional scattering theory. Also, because reflection occurs at a plane the negative reflection metamaterial is a patterned two-dimensional surface inherently far simpler than three-dimensional materials. It is reasoned that theses behaviors will lead to further discoveries and applications.

**Acknowledgements:**

**Figure Captions:**

1. Wave propagation at an interface between two media. An incoming wave vector is incident along direction AO; OB and OC mark the usual reflected and refracted wave directions, OC' and OB' are at negative angles and represent negative refraction and negative reflection respectively.

2. Non-specular scattering from a periodic scattering surface, here the incoming wave vector ($k_1$) defines the x-axis of a coordinate system fixed with respect to the laboratory. The heavy line (red) is the line of intersection between the scattering structure and the plane of the figure.

3. The $\alpha$, $\beta$ phase space include regions (shaded) of positive and negative reflections, rest of the plane is not allowed to reflection. Dependence of the possible values outgoing direction ($\beta$) as a function of the incident direction ($\alpha$) is shown. The conventional specular reflections lie along the diagonal $\beta = 2\alpha$ straight line. At the boundaries between the reflected and the transmitted regions critical diffraction takes place and the tow angles degenerate into $\beta = \alpha$ & $\alpha + \pi$. On both sides adjacent to these line non-specular scattering is very sensitive to changes in $\alpha$.

4. Experimental realization and demonstration of non-specular reflection in the visible regime. Left panel: two parallel beams of light (green and red) scatter from a periodic surface (shown in silhouette bottom left). The familiar (positive) reflection spots fall on the left however negatively reflected beams remain on the same side of the normal to the scattering surface. Right panel: details of spot position on the reflection the screen; the negative beams remain on the same side of the incident beams and for the relevant values of the experimental parameters more sensitive to wavelength.

5. Graphs of the calculated $\alpha$ and $\beta$ for six values of $\kappa$ (0.4, 0.6, 0.8, 1.0 & 1.2) and two sets of experimental data obtained for the red and green lasers are shown. Excellent agreement between experiment and theory is obtained. The graphs clearly display non-trivial dependency of $\beta$ and $\alpha$.

6. Heliostatic behavior. The calculated values of the grating constant required to keep a constant emergent angle vs. incident angle for six fixed values of $\beta$ are shown.

**Figure 1:**

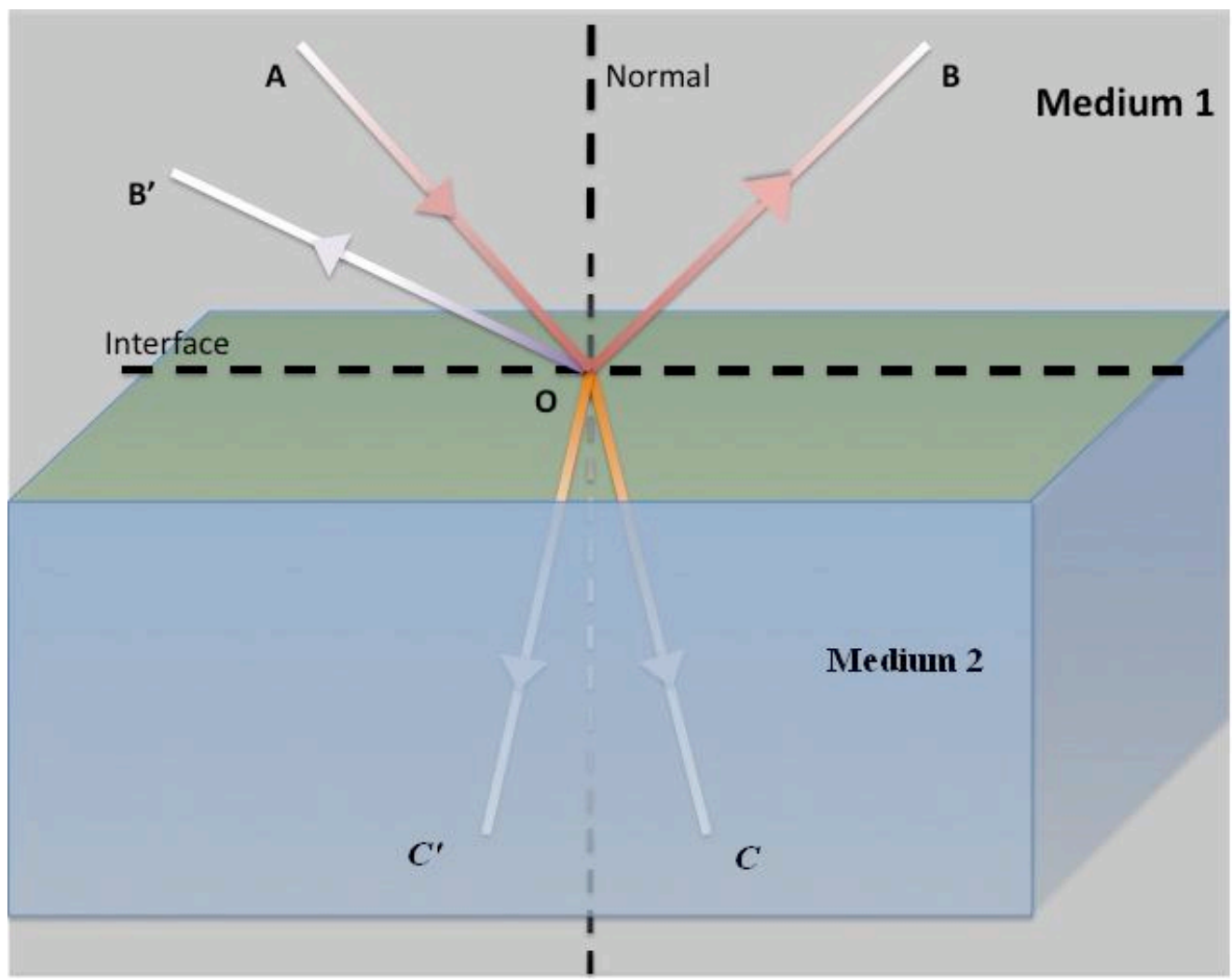

**Figure 2:**

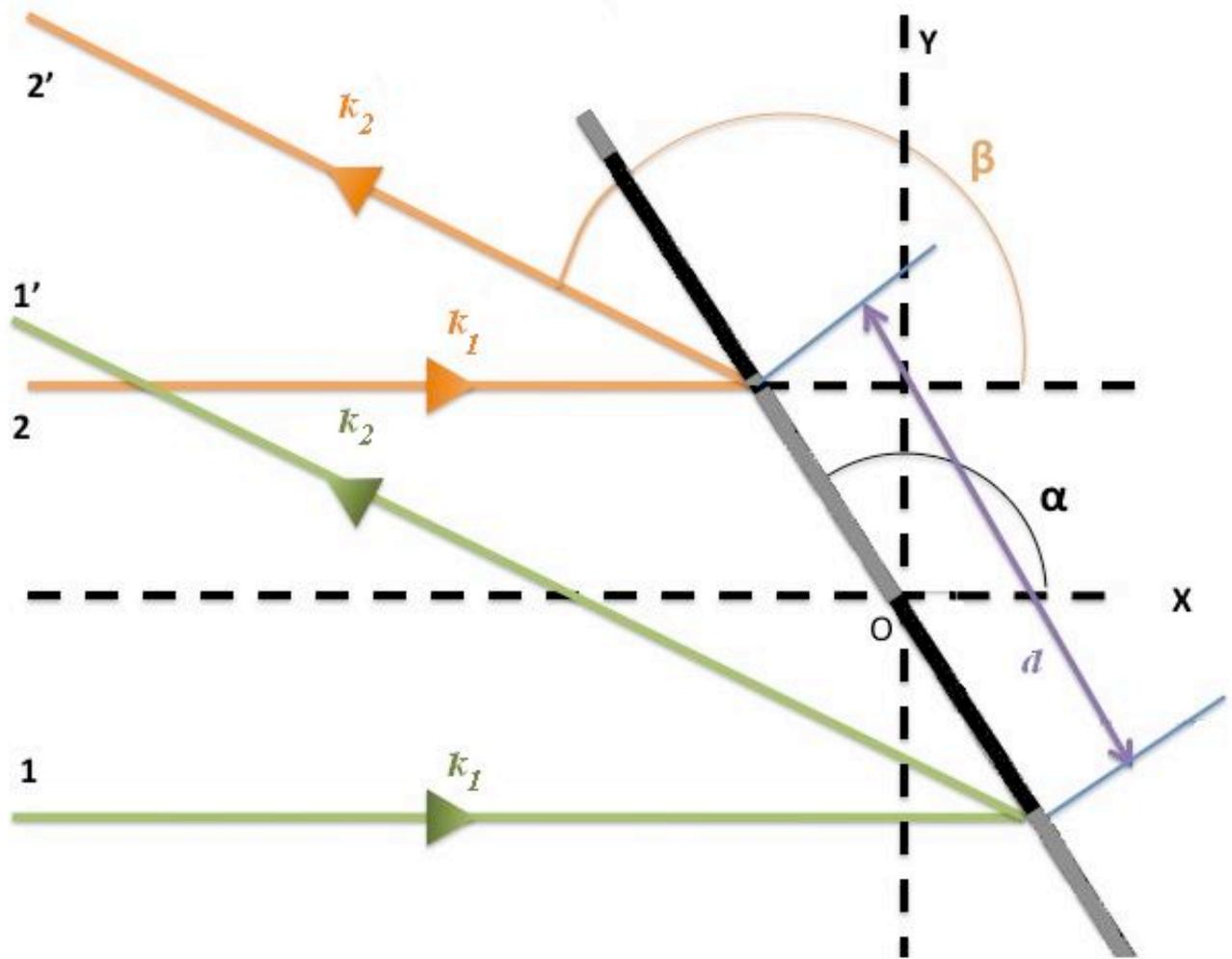

**Figure 3:**

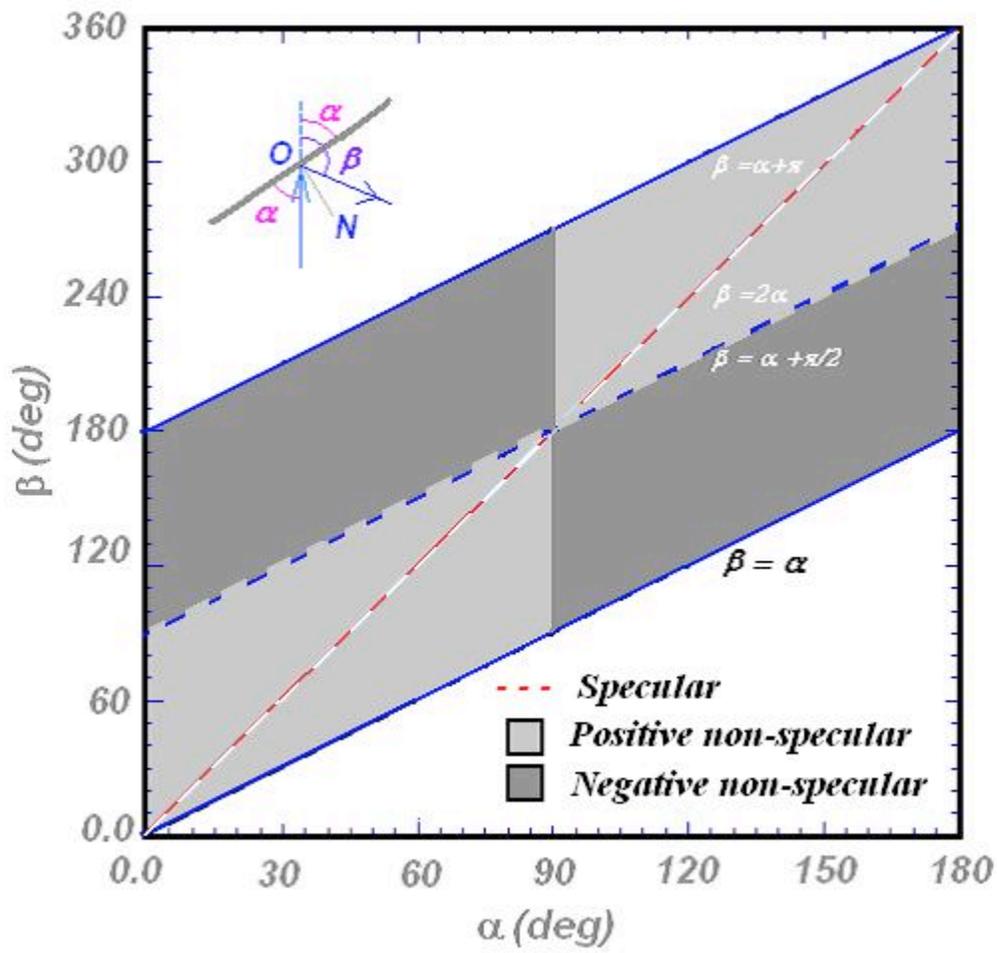

**Figure 4:**

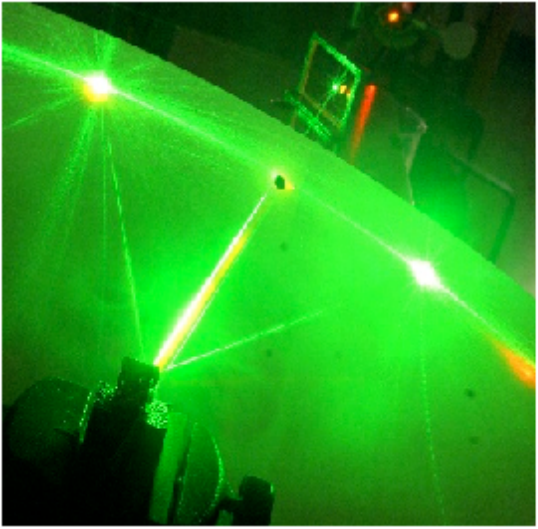 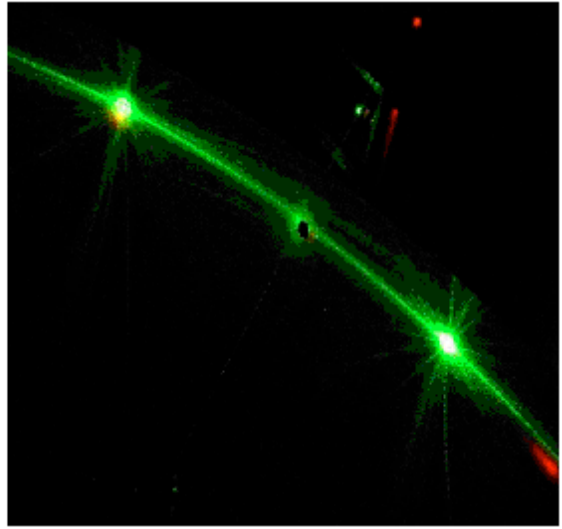

**Figure 5:**

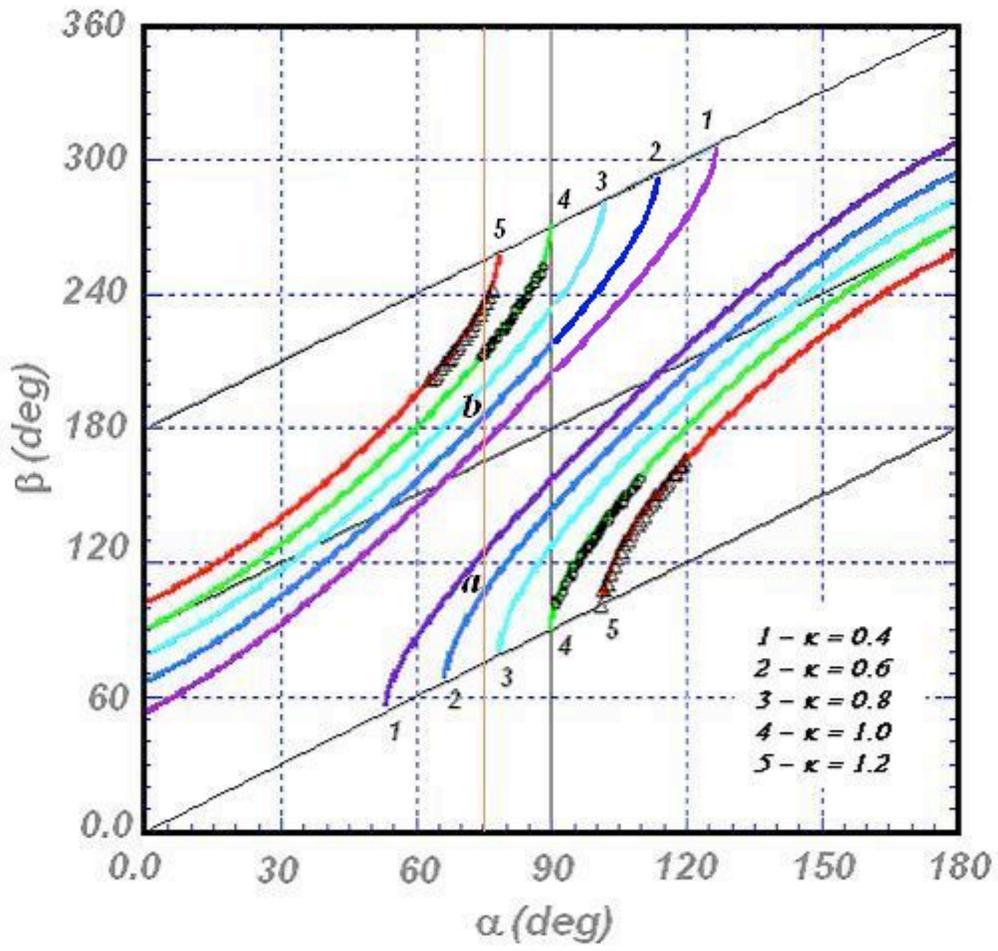

**Figure 6:**

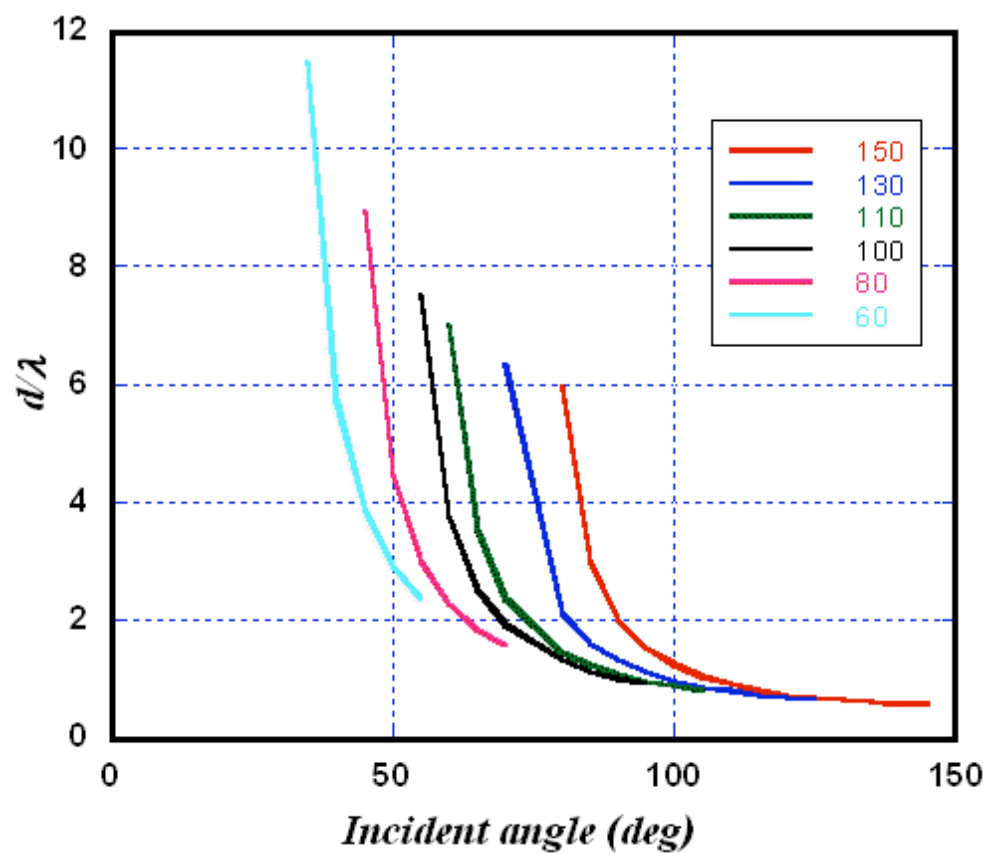